# *Cavity-stimulated Raman emission from a single quantum dot spin*

Timothy M. Sweeney[1], Samuel G. Carter[2], Allan S. Bracker[2], Mijin Kim[3], Chul Soo Kim[2], Lily Yang[1], Patrick Vora[1], Peter G. Brereton[4], Erin R. Cleveland[5], and Daniel Gammon[2]

[1] *NRC research associate at the Naval Research Laboratory, Washington, DC, 20375, USA*

[2] *Naval Research Laboratory, Washington, DC 20375, USA*

[3] *Sotera Defense Solutions, Inc., Annapolis Junction, MD 20701, USA*

[4] *US Naval Academy, Annapolis, MD 21402, USA*

[5] *ASEE research associate at the Naval Research Laboratory, Washington, DC, 20375, USA*

Solid state quantum emitters have shown strong potential for applications in quantum information, but spectral inhomogeneity of these emitters poses a significant challenge. We address this issue in a cavity-quantum dot system by demonstrating cavity-stimulated Raman spin flip emission. This process avoids populating the excited state of the emitter and generates a photon that is Raman shifted from the laser and enhanced by the cavity. The emission is spectrally narrow and tunable over a range of at least 125 GHz, which is two orders of magnitude greater than the natural linewidth. We obtain the regime in which the Raman emission is spin-dependent, which couples the photon to a long-lived electron spin qubit. This process can enable an efficient, tunable source of indistinguishable photons and deterministic entanglement of distant spin qubits in a photonic crystal quantum network.

Controlled absorption and emission of single photons by quantum emitters are essential processes for quantum information technologies. Single photons can be used to transfer quantum information from one stationary qubit to another as part of a quantum network[1–4], or they can be used as a qubit for photonic quantum computing[5] or secure communication[6]. Currently the largest challenge in achieving these goals is in scaling up the number of qubits. A promising approach is the integration of solid state quantum emitters into a photonic architecture[7–9]. Candidate materials include quantum dots[7,8] (QDs), QD molecules[10–13], nitrogen-vacancy centers in diamond[9,14], and other impurities or defects in solids[15,16]. These materials can take advantage of nanofabrication technologies to produce monolithic integrated structures that simplify the scaling-up problem[7,9]. Unfortunately, solid state quantum emitters suffer from spectral inhomogeneity, which greatly limits their usefulness for protocols that involve identical photons[5] or that involve the exchange of a photon between two qubits[1–4].

Here we demonstrate for the first time in a solid state system a cavity-stimulated Raman process[17–19] that can be used to overcome spectral inhomogeneity. We do this by coupling a negatively charged InAs/GaAs quantum dot (QD) that acts as a Λ-type quantum emitter to a photonic crystal defect cavity[20]. Most previous work on QDs in cavities involves the coupling of a 2-level exciton system to a cavity[7,21–23]. In contrast, the three level Λ-type system here provides a long-lived ground state electron spin coherence[24,25], with ultrafast optical gates[25–28] and cavity-enhanced initialization and readout[20]. The key feature of the Raman process (see Fig. 1a) is that the frequency of the emitted Raman photon is determined by the laser photon energy and the Zeeman energy, not by the excited state energy of the quantum emitter. The cavity strongly enhances this process when the Raman photon is resonant with the cavity mode. We measure cavity-stimulated Raman photons detuned from the QD over a range of at least 0.5 meV (125 GHz), which is much larger than the detuning that has been achieved without a cavity (4 GHz)[29,30], and the natural QD linewidth (~1 GHz without a cavity). Moreover, the Raman photons have a narrow bandwidth that is limited by the spin resonance linewidth rather than the QD optical transition linewidth[29]. We obtain the limit where the cavity-stimulated Raman process depends strongly on the spin state, which couples the spin qubit to the emission of a Raman photon from the cavity, essentially a photon number qubit. Thus, these results allow for the generation of indistinguishable photons from dissimilar quantum emitters and also provide a spin-photon interface, enabling a deterministic approach to controlled entanglement of two matter qubits in a photonic crystal.

**Spin-Cavity System**

These experiments are performed on a single InAs/GaAs QD coupled to a photonic crystal cavity. The photonic crystal membrane contains an n-i-p (n-type, intrinsic, p-type) diode structure[31–34] (Fig. 1c,d) to enable control of the charge state of the QD[20], as displayed in the photoluminescence bias map in Fig. 2a. We operate at a bias of 1.53-1.6 V where a single electron is stable within the QD, giving rise to emission from $X^-$, a negatively-charged exciton (trion). The L3 cavity mode is ~0.5 meV below $X^-$, small enough to observe cavity-stimulated Raman emission but large enough to avoid spectral overlap of the cavity with trion states. An in-plane magnetic field of 4 T induces Zeeman splittings according to the in-plane ground state electron g-factor (0.43) and excited state trion g factor (0.21), giving rise to four optical transitions. As illustrated in Fig. 2b, each trion state (labeled T1 and T2) forms a Λ-type three-level system with the two electron spin ground states. One transition (2 or 3) of each Λ-type system is well-coupled to the linearly-polarized cavity mode. The differential reflectance (DR) spectrum of the coupled cavity-QD system is plotted in Fig. 2c at a temperature of 5 K and with the probe laser polarized parallel to the PhC cavity. The cavity response is observed at 1290.7 meV with a quality factor (Q) of ~4,000 and has a dispersive lineshape due to interference effects and the differential nature of this technique. Optical transitions 2 and 3 are polarized nearly parallel to the cavity and are visible at 1291.2 meV in Fig. 2c. Optical transitions 2 and 3 have linewidths of ~18 μeV at this QD-cavity detuning, which are broadened due to coupling to the cavity. Optical transitions 1 and 4 are polarized nearly orthogonal to the cavity and are therefore very weak in Fig. 2c.

Typically, resonantly driving one of these transitions results in optical pumping[35,36]: with the laser tuned to one leg of the Λ system, recombination eventually takes it to the other leg, causing the driven transition to go dark. This occurs when the pumping rate is much higher than the spin relaxation rate ($T_1$ is 20 μs in this QD). For most of the experiments described here, we work at a bias on the edge of the charge stability range, where there are rapid spin flips due to cotunneling of electrons between the QD and the n-doped region of the diode[37]. This randomizes the spin state and prevents transitions from going dark, simplifying some measurements. Ultimately, a spin qubit requires long coherence times. Therefore at the end of this paper we present Raman experiments that exploit optical pumping to initialize the spin into a pure eigenstate.

**Cavity-stimulated Raman emission**

By spectrally resolving the emission in response to a drive laser, we can measure the spectrum of the coherent two-photon process known as spin-flip Raman emission as shown in Fig. 2d for a laser frequency near the cavity. The laser polarization is perpendicular to the cavity polarization, and the detection polarization is parallel. We use a scanning Fabry-Perot interferometer with a resolution of 1.7 μeV followed by a spectrometer with CCD detection. The emission spectrum shows two sidebands called Stokes (S) and anti-Stokes (AS) that are shifted from the laser (L) by the electron Zeeman energy ($E_z$). The photons emitted at these Raman-shifted frequencies are correlated with the electron spin-flip processes diagrammed in the insets of Fig. 2d. This spectrum is obtained at a laser frequency detuning from the QD of 0.5 meV – more than an order of magnitude larger than what was previously shown in the absence of a cavity[29,30]. This large frequency detuning is an important quality of cavity-stimulated Raman emission and so we characterize it in some detail— first, near the QD transitions with relatively weak laser power and then over the full cavity-QD spectral range at higher laser power.

In Fig. 3 we present the resonant Raman spectra in the region near the QD resonances at low laser power (100 nW). The DR spectrum of the QD transitions is shown again for comparison in Fig. 3a. The Raman spectra for a series of laser frequencies are plotted in Fig. 3b for both Stokes (top spectra) and anti-Stokes (bottom spectra). The intensities of the Raman peaks are plotted in Fig. 3c to give the Raman excitation spectra. When the laser is resonant with transition 1, the anti-Stokes Raman emission is resonant with the cavity-coupled transition 3 to give the strong resonant enhancement centered at transition 1. Likewise the Stokes excitation resonance is centered at transition 4. The linewidths of the Raman excitation resonances in Figs 3c are about the same as the DR resonances (18 μeV) shown in Fig. 3a.

We note that on resonance with the QD transitions the photon produced in the spin-flip process could either be created directly in the Raman process, or alternatively, in sequential processes where the laser photon is first absorbed and then incoherently emitted. These two processes can be differentiated by measuring the emission linewidth and tuning behavior. Raman emission tunes with the excitation laser and has the linewidth of the spin dephasing rate, whereas absorption followed by incoherent emission will emit at, and have the linewidth of, the optical transition[29]. The Raman emission sidebands in Fig. 3b track the laser and are much narrower (3 µeV) than the cavity-broadened optical linewidths of 18 µeV. In fact we find this to be the case at all laser detunings. Therefore we conclude that the Raman process dominates, as is expected to occur when pure optical dephasing is weak[38]. The linewidth of the Raman sidebands corresponds to a spin dephasing time of ~0.5 ns, which is consistent with previous measurements of electron spins in these QDs interacting with nuclear spins[39]. The spin lifetime can be extended to much longer times by minimizing the effects of nuclear spin[24,25]. Spin relaxation due to co-tunneling can also contribute, but we find no significant difference in the Raman emission linewidth when co-tunneling is inhibited. In contrast to incoherent PL, the Raman emission line also tunes with the drive laser and is observed over a much wider range than the QD linewidth.

In Fig. 4a we again plot the amplitudes of the Raman emission sidebands as a function of laser frequency, but now over a wider frequency range and at higher laser power (15 µW). We observe a highly asymmetric response as a function of detuning from the QD for both Stokes and anti-Stokes processes because of a strong enhancement by the cavity. This is a clear indication of cavity-stimulated Raman emission. The asymmetry is further illustrated in Fig. 4b, where the Raman emission spectrum is plotted at laser detunings of +/-440 ueV from the QD. The Raman signal when the laser is near resonant with the cavity (-440 µeV) is ~20 times stronger than the Raman signal when the laser is blue-detuned from the dot (+440 µeV). The Raman emission amplitude can be modeled in second-order perturbation theory[38], modified by the cavity-enhanced density of states for the Raman emission. The result is the product of two Lorentzians (taken from the experimental linewidths), one with the laser centered on QD transitions 1 or 4, and one centered on the cavity and shifted by the Zeeman energy (see Methods). Here, the wide tuning range is demonstrated with the Raman photons tuned across the somewhat broad cavity resonance. Better performance could be achieved with a higher Q cavity that is tuned[40,41] *in situ* to match a chosen Raman photon energy.

In Fig. 4c we display the second order correlation function $g^{(2)}(\tau)$ of the Raman photons, obtained with the QD lines tuned within the cavity linewidth (~150 µeV above the cavity resonance). As expected for a single photon source, there is a pronounced antibunching dip at $\tau = 0$ that is well-fit by an exponential rise, $1-\exp(-|\tau|/t_{rise})$ with $t_{rise}$ = 1.1 ns, convoluted with the temporal response of the detectors.

**Spin selectivity and correlation between Raman intensity and spin state**

We now demonstrate that the cavity-stimulated Raman process can be made spin-selective. This is necessary, for example, to map the QD spin state onto an emitted photon[1]. When the laser is near resonance with the cavity and detuned from the QD, as shown in Fig. 4b, both strong Stokes and anti-Stokes lines are observed, in contrast to when the laser is near resonance with the QD in

Fig. 3. This is due to the Raman response for both processes being within the bandwidth of the cavity ($2\Gamma=350$ µeV). However, we can select one process over the other if the laser and cavity frequencies meet the Raman resonance condition for one transition but not the other (see Fig. 5a). This requires that the energy difference between the Stokes and anti-Stokes frequencies ($2E_z$) be larger than the cavity half-width ($\Gamma$). We show the selectivity $(I_{AS} - I_S)/(I_{AS} + I_S)$ as a function of magnetic field in Fig. 5b. For this experiment the laser frequency is tuned at each magnetic field value to keep the anti-Stokes emission frequency at the cavity peak. As a result the Stokes emission frequency shifts away from the cavity peak and the amplitude decays accordingly. An example of the Raman spectrum under these conditions at B = 4 T is shown in Fig. 5c, again with an unpolarized spin. The anti-Stokes peak is seven times that of the Stokes. As shown in Fig. 5b the measured selectivity as a function of magnetic field is well fit using the cavity linewidth and no adjustable parameters (see Methods). Higher fields or sharper cavity linewidths will lead to larger selectivity.

The results presented in this paper thus far characterize the cavity-stimulated Raman emission process using randomized electron spin. We now prepare the spin into a pure eigenstate to show that Raman emission is correlated with the spin state of an electron. This is essential for reliable on-demand cavity-stimulated Raman photons or for using spins as qubits in a distributed quantum network. We accomplish this by tuning the voltage bias to the center of the stability plateau such that spin relaxation due to co-tunneling is small relative to the optical pumping rate. The addition of a second laser in this experiment is used to initialize the spin state to either down or up through optical pumping of the appropriate QD transition. To initialize the QD into the spin up (down) state we tune the second laser to optical transition 1 (4) shown in figure 3a. Under these conditions the anti-Stokes emission at the cavity resonance depends strongly on whether the spin is up or down while the Stokes is suppressed as shown in Figs. 5d and 5e. This result explicitly shows that the cavity-stimulated emission is correlated with the spin state. It should also be possible to map a coherent spin superposition state onto a Raman photon number state in this way, although this has not yet been done.

**Discussion/Outlook**

We have demonstrated cavity-stimulated Raman emission in a solid state qubit-microcavity system. This result could enable a triggered single photon source with a tunable, sharp emission line. Due to the coherent nature of this process, which avoids populating the trion state, the Raman photons will be emitted with precise timing with respect to the drive laser and enable control of the photon temporal shape[42]. Using drive laser pulses shorter than the spin dephasing time will thus produce single photons with short temporal duration, suppressing effects from nuclear spin fluctuations and producing identical photons. The rate at which these single photons can be generated is ultimately limited by spin-initialization and cavity-stimulated Raman emission rates, which should enable GHz rate emission. Currently the Raman photon count rate is limited by detection, collection loss and poor coupling to free space, but efficient coupling to a waveguide [43] should enable much higher efficiencies.

This result is also quite significant for using QD spins for quantum computing. Perhaps the biggest fundamental hurdle in scaling up the number of qubits arises from the large distribution in frequencies. This is true of all solid state emitters, but especially with self-assembled QDs. While there has been considerable progress in optically measuring and controlling single pairs of QDs through coherent tunneling[11,12], scaling to many qubits will likely require coupling through photons. There has been much recent progress in this area as researchers have demonstrated entanglement between a QD spin qubit and an emitted photon[44–46] and even entanglement of two NV centers by interfering two photons that are each entangled with separate spin qubits[47]. Interference of indistinguishable photons has been achieved from two QDs[30,48], in one case using a spin-flip Raman photon[30]. This requires the emitters to be very closely matched in frequency, which is a significant issue for solid state emitters.

This demonstration of cavity-stimulated Raman emission greatly improves the prospects of photon-mediated entanglement. First, the cavity increases the emission rate of photons and can efficiently direct them to another cavity-QD system[49]. Second, as we have demonstrated here, the cavity permits tuning the photon frequency over a range much larger than the natural QD linewidth. Thus, QDs can have different transition energies because only the Raman photons need to be matched, and these are controlled by the drive laser. Likewise, any differences in the electron Zeeman energy between QDs can be compensated by tuning the individual laser frequencies for each QD. Here, we have demonstrated a tuning range of 500 μeV, but it is likely this can be extended with higher quality-factor cavities. Thus the combination of large tunability, high emission rates, and sharp linewidths make cavity-stimulated Raman emission a potentially important ingredient for solid state quantum networks.

**Methods**

**Sample.** The sample was grown by molecular beam epitaxy with the InAs QDs grown at the center of an n-i-p diode, with the layer thicknesses shown in Fig. 1d. The QDs were distributed randomly in the growth plane with a density of several QDs per $\mu m^2$. The two-dimensional photonic crystal consists of a triangular lattice of holes (70 nm radius) with a lattice constant of 244 nm that are etched through the diode epilayer into an $Al_{0.7}Ga_{0.3}As$ sacrificial layer. Three missing holes at the center form an L3 cavity (Fig. 1c). This pattern is defined by electron-beam lithography and a $Cl_2$-based inductively coupled plasma etch. The AlGaAs is removed under each photonic crystal using a selective etch, leaving a 180-nm-thick photonic crystal membrane. Ohmic contacts to the diode enable charge control of QDs.

Originally the QD-cavity detuning was 1.8 meV, with a QD linewidth of 6 μeV. The QD-cavity detuning was set to 0.5 meV by depositing ~2 monolayers of $Al_2O_3$ via atomic layer deposition[50], which red-shifted the cavity mode closer to the QD. Temperature tuning was not used in this system because the electron spin relaxation time ($T_1$) decreases from 20 µs at 5.2 K to 70 ns at 16 K.

**Raman emission model**. The Raman emission can be modeled in second-order perturbation theory[38], modified by the cavity-enhanced density of states for the Raman emission. The intensities of the Stokes and Anti-Stokes emission are then given by:

$$I_S \propto \frac{\mu_{T1,\uparrow}^2 \mu_{T1,\downarrow}^2}{(\omega_L - \omega_{T1,\uparrow})^2 + \gamma^2} D(\omega_L - \omega_z), \quad I_{AS} \propto \frac{\mu_{T2,\downarrow}^2 \mu_{T2,\uparrow}^2}{(\omega_L - \omega_{T2,\downarrow})^2 + \gamma^2} D(\omega_L + \omega_z),$$

where $\mu_{a,b}$ is the dipole moment between states $a$ and $b$, $\omega_{a,b}$ is the frequency difference between states $a$ and $b$, $\omega_L$ is the laser frequency, and $\gamma$ is the QD transition linewidth ($2\gamma = 18$ µeV). The cavity photon density of states $D(\omega)$ is given by $D(\omega) = \Gamma^2 / [(\omega - \omega_c)^2 + \Gamma^2]$, where $2\Gamma$ is the full-width-half-maximum of the cavity ($2\Gamma = 350$ µeV) and $\omega_c$ is the cavity frequency. In this perturbative model, the spin populations are assumed to be equal and saturation is ignored. These assumptions may not be valid for the strong, resonant excitation in Fig. 4a, which may account for discrepancies in the agreement with the data near resonance with the quantum dot. The selectivity plotted in Fig. 5b is given by substituting the above expressions for $I_S$ and $I_{AS}$ into the expression for selectivity $(I_{AS} - I_S)/(I_{AS} + I_S)$ with the anti-Stokes photon frequency centered on the cavity $\omega_{AS} = \omega_L + \omega_z = \omega_c$.

**Measurement techniques.** The DR spectrum was obtained utilizing phase sensitive detection, in which the sample bias is modulated at 1-5 kHz with an amplitude of 400 mV, modulating the charge state of the QD. The change in reflection due to this modulation is measured as the frequency of a diode laser is scanned. For measurements of Raman emission we employed a two stage system to spectrally resolve emission with high resolution. In the first stage we use a home-built scanning plano-concave Fabry-Perot interferometer with a resolution of 1.7 µeV and free spectral range of 400 µeV. To distinguish between different orders of the interferometer, the output is sent to a 750 mm spectrometer with a 1200 lines/mm grating and CCD detection.

The second order correlation function $g^{(2)}(\tau)$ of the Raman photons displayed in Fig. 4(c) was obtained by sending photons to a Hanbury Brown Twiss interferometer consisting of a beamsplitter and two silicon photon counting modules. The detectors have a time resolution of ~300 ps. A Gaussian convolution function $\exp(-\tau^2 / \Delta t^2)$ with $\Delta t = 400$ ps was used to simulate the effect of the response time on $g^{(2)}(\tau)$. Raman photons were isolated from reflected laser light through polarization rejection and the spectrometer without using the Fabry-Perot interferometer. For 1 µW excitation there were ~6,000 counts/s on each detector, and coincidence count statistics were built up over 17 hours.

 **Acknowledgments**

The authors would like to thank John Lawall for advice in high resolution spectroscopy and in building a scanning Fabry-Perot filter. The authors would also like to thank Hailin Wang for fruitful discussion into the physics of cavity-QED.

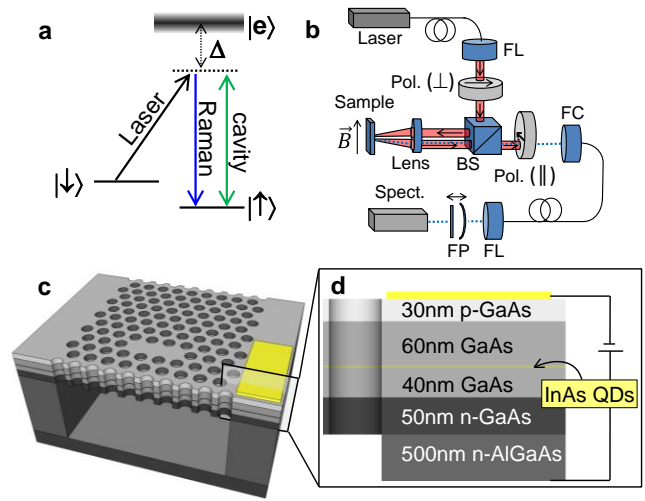

**Figure 1 | Concept: Raman emission from a cavity-QD system. a**, Λ-type energy level diagram illustrates Raman coupling between two spin states with laser and cavity frequencies at equal detunings from the excited state. **b**, Schematic diagram of the Raman emission experimental setup, with elements abbreviated as follows: scanning Fabry-Perot cavity (FP), polarizers (Pol.), beam splitter (BS), fiber launchers/collector (FL/FC), and spectrometer (Spect.). The input polarizer is set perpendicular (⊥) and detection parallel (∥) to the sample cavity mode. **c**, Illustration of an L3-type photonic crystal cavity etched into a membrane, with **d**, layer thicknesses of the n-i-p diode, showing the InAs QDs at the center.

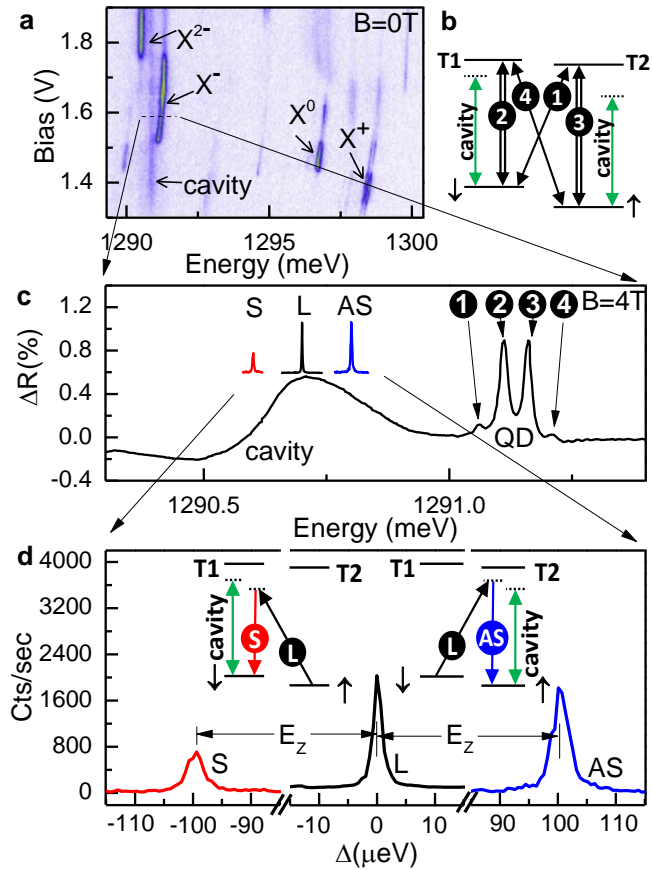

**Figure 2 | InAs quantum dot coupled to an optical cavity**. **a**, Photoluminescence bias map indicating controlled QD charging as a function of applied voltage. **b**, Energy level diagram of the charged quantum dot embedded in a linearly polarized cavity. The cavity mode is polarized nearly parallel to transitions 2 and 3 (indicated by the double lines) and nearly orthogonal to 1 and 4. **c**, ΔR scan with the laser polarized parallel to the cavity, showing the cavity resonance at 1290.7 meV and the 4 optical transitions near 1291.2 meV. Inset is the Raman response when the laser is near resonant but orthogonal to the cavity polarization. **d** Energy scalesare expanded for Stokes (S) and anti-Stokes (AS) emission from the QD each shifted by 100 μeV from the pump laser (L) that is polarized orthogonal to the cavity. Inset diagrams illustrate the AS and S processes.

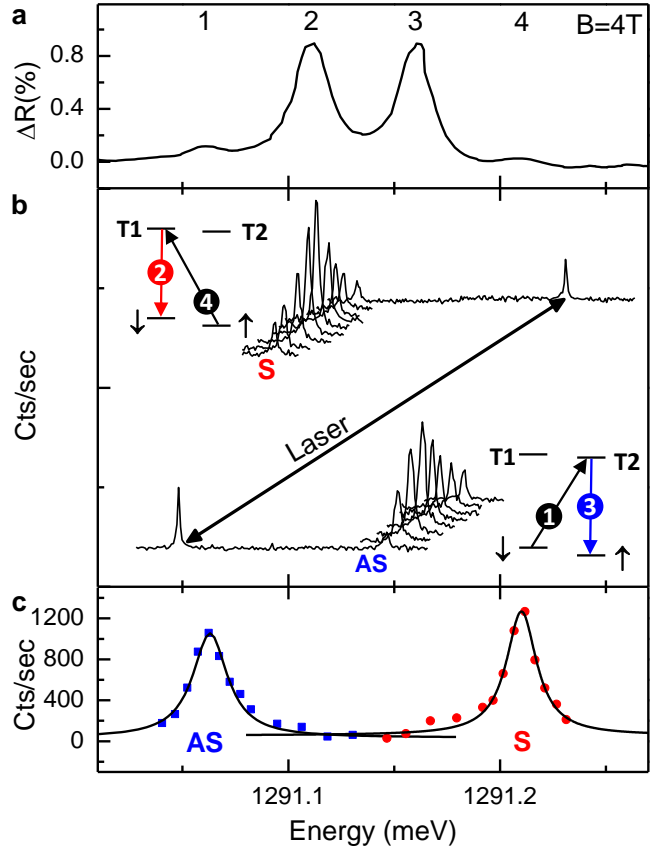

**Figure 3 | Resonant Raman emission**. **a**, The differential reflectance (ΔR) spectrum from the QD with the laser polarized parallel to the cavity and nearly parallel to transitions 2 and 3 at a magnetic field of B = 4 T. **b,** Examples of the Raman emission spectra (labelled S and AS) as the laser is tuned through transitions 1 and 4. The 100 nW laser is polarized perpendicular to the cavity with detection parallel. The level diagrams in **b** illustrate the AS and S processes. **c**, The Raman emission intensities for both the anti-Stokes (blue squares) and the Stokes (red circles) spectra in **b** are plotted to give the Raman excitation spectra. Lorentzian fits are also plotted, giving peak frequencies and FWHM (18 μeV) in agreement with the ΔR spectrum in **a**.

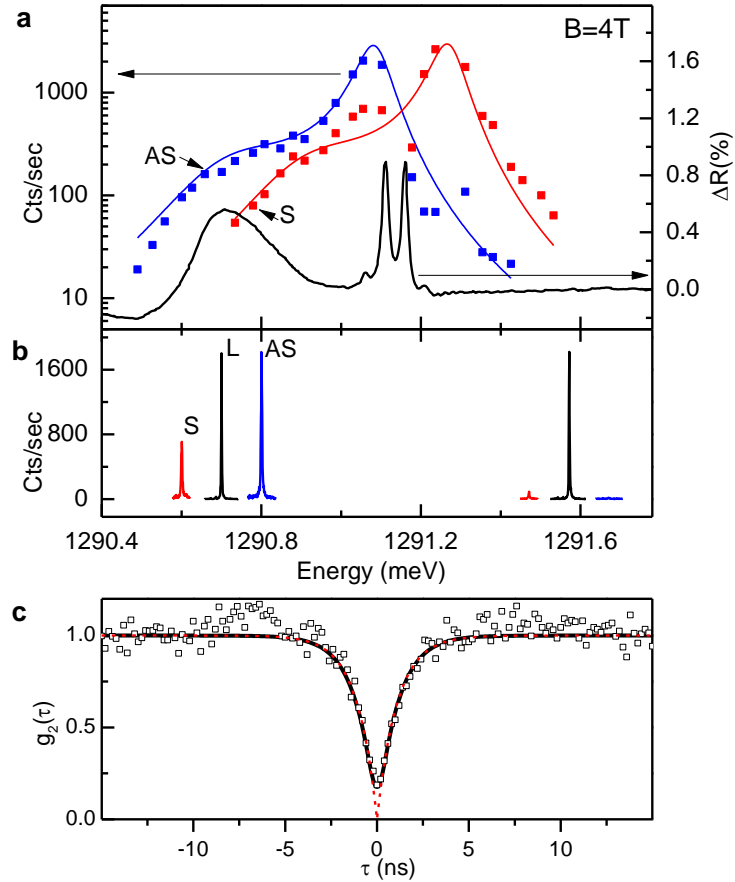

**Figure 4 | Cavity-stimulated Raman emission. a**, Raman emission amplitude as a function of laser frequency for a laser power of 15 µW. The red and blue lines are fits to the spectra based on the cavity-stimulated Raman process. The ΔR spectrum of the cavity-QD system is displayed for comparison. **b**, Two example Raman emission spectra at a laser power of 40 µW, each with similar detuning from the QD. The Raman signal from the cavity side of the QD is enhanced by a factor of ~20 compared to the higher energy side of the QD. **c**, Second order correlation function of Raman photons taken at 6.75 T for a laser power of 1 µW, with the laser resonant with transition 4 and the Stokes photon resonant with the cavity. The dashed line is an exponential rise function with a time constant of 1.1 ns, and the solid line convolutes this function with a Gaussian.

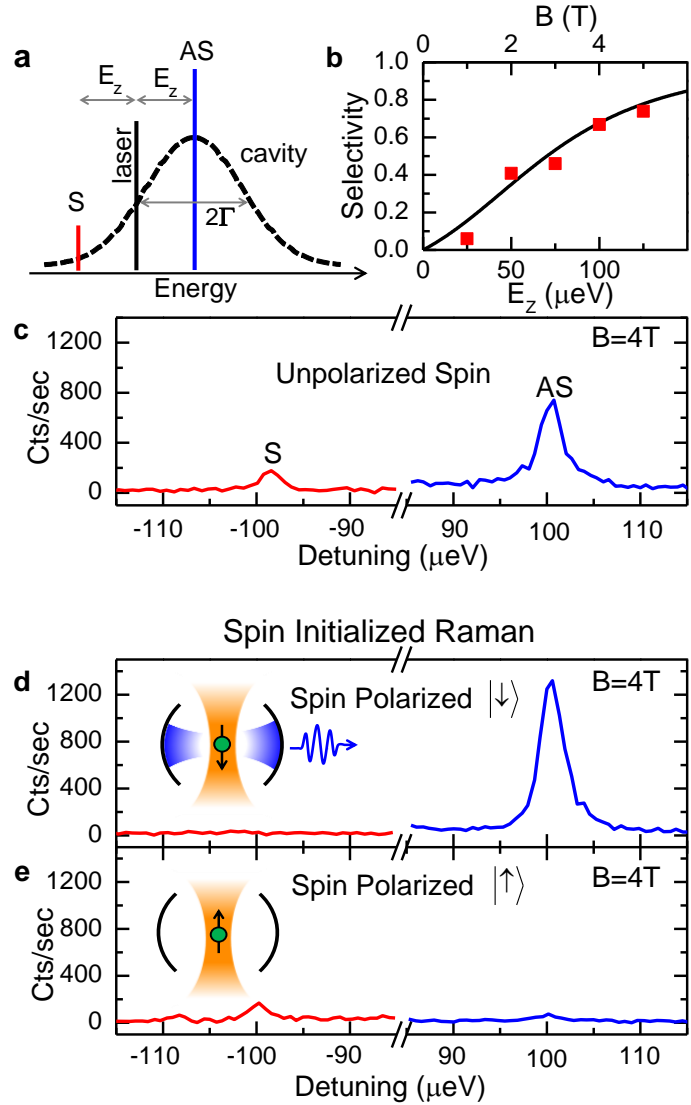

**Figure 5 | Spin selectivity and spin-photon correlation. a**, Diagram of cavity stimulated Raman selectivity. The AS line is centered at the cavity frequency and the laser is detuned from the cavity resonance by E$_z$. The diagram illustrates AS is more likely than S when $2E_z > \Gamma$. **b**, Measured Raman spin selectivity of AS over S as a function of $E_z$. The black line is the calculated selectivity using the model discussed in Methods. **c-e,** Raman emission spectra with **c**, unpolarized spin, **d**, spin polarized down, and **e**, spin polarized up. Insets in **d** and **e** illustrate spin selective cavity-stimulated Raman. The laser (orange) produces AS (blue) emission for spin down **d** and no Raman emission for spin up **e**.

16